\documentclass[twocolumn,floatfix,superscriptaddress,showpacs,showkeys]{revtex4}
\usepackage{graphicx}
\usepackage{amsmath}
\usepackage{booktabs}
\usepackage{amssymb}
\usepackage{multirow}
\usepackage{makecell}
\usepackage{textcomp}
\usepackage{subfigure}
\usepackage{phonetic}
\usepackage{extarrows}
\usepackage{color}
\usepackage{float}
\usepackage[colorlinks,citecolor=blue,linkcolor=blue]{hyperref}
\begin{document}
\title{Distinguishing photon blockade in a $\mathcal{PT}$-symmetric optomechanical system}
\author{Dong-Yang Wang}
\affiliation{Department of Physics, Harbin Institute of Technology, Harbin, Heilongjiang 150001, China}
\author{Cheng-Hua Bai}
\affiliation{Department of Physics, Harbin Institute of Technology, Harbin, Heilongjiang 150001, China}
\author{Shutian Liu\footnote{E-mail: stliu@hit.edu.cn}}
\affiliation{Department of Physics, Harbin Institute of Technology, Harbin, Heilongjiang 150001, China}
\author{Shou Zhang\footnote{E-mail: szhang@ybu.edu.cn}}
\affiliation{Department of Physics, College of Science, Yanbian University, Yanji, Jilin 133002, China}
\author{Hong-Fu Wang\footnote{E-mail: hfwang@ybu.edu.cn}}
\affiliation{Department of Physics, College of Science, Yanbian University, Yanji, Jilin 133002, China}
\begin{abstract}
We study the effects of parity-time($\mathcal{PT}$)-symmetry on the photon blockade and distinguish the different blockade mechanisms in a double-cavity optomechanical system. By studying the light statistics of the system, we find the completely different photon blockade behaviors when the $\mathcal{PT}$-symmetry is broken or unbroken, which is related to the $\mathcal{PT}$ phase transition. Furthermore, an interesting phenomenon that the two cavities are blocked at the same time is found with the appropriate system parameters. Those statistical phenomenons are all analyzed in detail and demonstrated by analytically solving the Schr\"{o}dinger equation and numerically simulating the master equation, respectively. Finally, we also consider the non-$\mathcal{PT}$ symmetric situations which further reveal the physical essence of the photon blockade by comparing those results. Different from the usual photon blockade, our proposal is feasible even with weak parameter mechanism, i.e., the proposal neither requires the strong optomechanical coupling nor the large tunneling coupling between cavities.
\end{abstract}
\maketitle

\section{Introduction}\label{sec.1}
Cavity optomechanics~\cite{RevModPhys.86.1391,andp.525.215,Science.321.1172,PhysToday.65.29,Physics.2.40} has made great success in the exploration of macroscopic quantum effects and the study of interaction between electromagnetic radiation and micromechanical motion over the past decades. Various cavity optomechanical systems~\cite{PhysRevLett.110.193602,NatComm.6.6981,OptLett.41.2422,PhysRevA.90.043831,NatComm.7.11338,arXiv:1810.03709} have been proposed and become the remarkable platforms to study the questions of quantum mechanics on the macroscopic scale, such as the ground-state cooling~\cite{Nature.524.325,PhysRevLett.99.093901,PhysRevLett.99.093902,Nature.463.72,Nature.444.67,PhysRevA.98.023816}, mechanical quadrature squeezing~\cite{Science.349.952,PhysRevX.3.031012,PhysRevLett.103.213603,PhysRevA.88.013835}, macroscopic entanglement~\cite{PhysRevLett.98.030405,PhysRevLett.99.250401,PhysRevLett.88.120401,arXiv:1811.06227}, quantum superposition state of the mechanical resonator~\cite{PhysRevLett.116.163602,PhysRevLett.110.160403,PhysRevA.88.023817}, etc. In addition, the electromagnetic field (optical part) of the cavity optomechanical system is also affected by the motion of the mechanical resonator. Some researches have attracted significant attention, such as the optomechanically induced transparency~\cite{PhysRevA.81.041803,Science.330.1520,Nature.472.69}, normal-mode splitting~\cite{PhysRevLett.101.263602}, photon blockade~\cite{PhysRevLett.107.063601,PhysRevLett.107.063602,PhysRevA.87.025803,PhysRevA.88.023853,PhysRevA.92.033806,PhysRevA.93.063860,arXiv:1802.09254}, etc. Of these studies, the photon blockade is a particularly important nonclassical light statistic effect and can be used to generate the single photon source for those fundamental studies in quantum information processing and quantum optics fields.

To this end, there are two general ideas: (i) the anharmonicity of eigenenergy spectrum coming from kinds of nonlinearities, which is called as the conventional photon blockade (CPB); (ii) the destructive quantum interference between two different transition paths, which is called as the unconventional photon blockade (UPB). The first physical mechanism for achieving blockade has been studied theoretically~\cite{PhysRevLett.79.1467,PhysRevA.49.R20,PhysRevA.46.R6801,PhysRevB.87.235319,PhysRevA.90.023849} and realized experimentally in various systems~\cite{Nature.436.87,PhysRevLett.107.053602,NatPhys.4.859}. Besides, due to the nonlinear optomechanical interaction, the optomechanical systems can also be used to achieve the photon blockade effect~\cite{PhysRevLett.107.063601,PhysRevLett.107.063602,PhysRevA.87.025803,PhysRevA.88.023853,PhysRevA.92.033806,PhysRevA.93.063860,arXiv:1802.09254}. However, it is worth noting that the study of photon blockade in an optomechanical system requires the strong optomechanical coupling condition, which still poses major technological challenges. On the other hand, the physical mechanism of the destructive quantum interference for achieving blockade has been proposed~\cite{PhysRevLett.104.183601} and studied extensively in previous years~\cite{PhysRevA.83.021802,PhysRevA.90.033809,PhysRevA.91.063808,PhysRevA.92.023838,PhysRevA.92.053837,OE.23.32835,JPB.51.035503,PhysRevA.96.053810,PhysRevA.97.043819,PhysRevA.98.023856}. Recently, it has been observed experimentally in coupled quantum-dot-cavity system~\cite{PhysRevLett.121.043601}. Based on the theory, the UPB has also been studied in various double-cavity optomechanical systems~\cite{PhysRevLett.109.013603,PhysRevA.87.013839,JPB.46.035502,PhysRevA.98.013826}. It is worth noting that the UPB usually requires the large coupling strength between the two components~\cite{PhysRevA.96.053810}, which structure the different transition paths. In a word, the usual photon blockade requires strong parameter mechanism, i.e., large nonlinearity or strong coupling in different blockade mechanism. So how to break the limitation of strong parameter mechanism will be beneficial to the experimental realization.

Moreover, the $\mathcal{PT}$-symmetric system exists plenty of interesting physical behaviors~\cite{PhysRevLett.80.5243,RepProgPhys.70.947} and has been studied in various physical fields~\cite{PhysRevLett.103.093902,PhysRevLett.103.123601,PhysRevLett.106.213901,Nature.488.167,NatMat.12.108,NatPhys.6.192,PhysRevA.96.043810,Science.346.975,NatPhotonics.8.524,Nature.548.187,Science.363.eaar7709}. Ref.~\cite{PhysRevA.92.053837} has studied the enhancement of photon blockade in a $\mathcal{PT}$-symmetric coupled-cavity system. Naturally, the study of the $\mathcal{PT}$-symmetry in cavity optomechanics has also attracted much attention in recent years, such as the phonon laser~\cite{PhysRevLett.113.053604}, chaos~\cite{PhysRevLett.114.253601}, optomechanically induced transparency~\cite{SR.5.9663,SR.6.31095,PhysRevA.98.033832}, metrology~\cite{PhysRevLett.117.110802}, etc. Of these proposals, the $\mathcal{PT}$-symmetric optomechanical systems usually consist of two optical cavities and one mechanical mode; namely, the linear double-cavity optomechanical system, where the passive optical cavity is coupled to the mechanical resonator. In this way, studying the effects of $\mathcal{PT}$-symmetry on the photon blockade in optomechanical system will be interesting and important, and has not yet been reported so far.

In this paper, we focus on studying the effects of $\mathcal{PT}$-symmetry on the photon blockade in a double-cavity optomechanical system, where the optomechanical cavity is passive and the other one is an active cavity. We obtain the equal-time second-order correlation functions of the photon via analytically solving the Schr\"{o}dinger equation or numerically simulating the master equation, where the analytical and numerical results agree with each other very well. We find that the two cavities can be blocked simultaneously with appropriate system parameters and the photon blockade behaviors are completely different in the different $\mathcal{PT}$ phase regions, i.e., the correlation has three dips in the unbroken $\mathcal{PT}$-symmetric region, while in the broken $\mathcal{PT}$-symmetric region, there is only one dip. To explore and distinguish the deeply physical mechanisms of the different photon blockade behaviors, we analyze the $\mathcal{PT}$-symmetric double-cavity optomechanical system by utilizing the theories of CPB and UPB, respectively. We find that, in the unbroken $\mathcal{PT}$-symmetric region, the three dips come from the anharmonicity of eigenenergy spectrum and the destructive quantum interference between paths, which correspond to the CPB and UPB, respectively. However, in the broken $\mathcal{PT}$-symmetric region, the only one dip belongs to the UPB, which comes from the destructive quantum interference between paths. In addition, the present UPB is slightly different from the usual theory, where the interference paths are not affected by the $\mathcal{PT}$-symmetry. Finally, we also take into account the situations of non-$\mathcal{PT}$ symmetry: (i) double passive cavity; (ii) unequal detunings (different cavity resonance frequencies). We find that the balance gain-loss ratio is the main reason for the perfect photon blockade and the different detunings just change the location of the perfect photon blockade. Furthermore, the perfect photon blockade would not be achieved under the weak parameter mechanism if the balance is broken.

The paper is organized as follows: In Sec.~\ref{sec.2}, we derive the Hamiltonian of the $\mathcal{PT}$-symmetric double-cavity optomechanical system. In Sec.~\ref{sec.3}, we analytically and numerically solve the equal-time second-order correlation functions and analyze the effects of $\mathcal{PT}$ phase transition on the photon blockade. In Sec.~\ref{sec.4}, we discuss the photon blockade behavior when the system is non-$\mathcal{PT}$ symmetric. Finally, a conclusion is given in Sec.~\ref{sec.5}.

\section{System and Hamiltonian}\label{sec.2}
\begin{figure}
	\centering
	\includegraphics[width=0.65\linewidth]{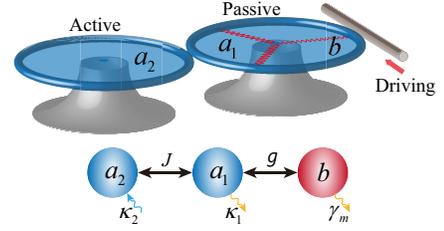}
	\caption{Schematic diagram of the $\mathcal{PT}$-symmetric double-cavity optomechanical system comprised of the whispering-gallery resonators.}
	\label{fig:double cavity optomechangcs}
\end{figure}

Inspired by the studies about the $\mathcal{PT}$-symmetric optomechanical systems~\cite{PhysRevLett.113.053604,PhysRevLett.114.253601,SR.5.9663,SR.6.31095,PhysRevA.98.033832,PhysRevLett.117.110802}, we consider a double-cavity optomechanical system, as depicted in Fig.~\ref{fig:double cavity optomechangcs}, where one optical cavity is passive and coupled to the mechanical resonator with single photon optomechanical coupling strength $g$, and the other one is an active cavity coupled to the passive cavity with tunneling coupling strength $J$, which can be controlled by changing the distance between the whispering-gallery resonators~\cite{Nat.Phys.10.394}. In the presence of the external driving field, the total Hamiltonian for the system is written as ($\hbar=1$)
\begin{eqnarray}\label{e01}
H&=&\omega_{1}a_{1}^{\dagger}a_{1}+\omega_{2}a_{2}^{\dagger}a_{2}
	+\omega_{m}b^{\dagger}b+J(a_{1}^{\dagger}a_{2}+a_{1}a_{2}^{\dagger})\cr\cr
	&&-ga_{1}^{\dagger}a_{1}(b^{\dagger}+b)+(Ea_{1}^{\dagger}e^{-i\omega_{l}t}+E^{\ast}a_{1}e^{i\omega_{l}t}),
\end{eqnarray}
where $a_{j}~(b)$ and $a_{j}^{\dagger}~(b^{\dagger})$ represent the annihilation and creation operators for the $j$-th optical (mechanical) mode with frequency $\omega_{j}~(\omega_{m})$, respectively ($j=1,2$). $E=\sqrt{2\kappa_{1}P/(\hbar\omega_{l})}$ is the driving amplitude of the input laser with frequency $\omega_{l}$ and power $P$, and $\kappa_{1}$ is the decay rate of the passive cavity 1. The gain rate of the active cavity (cavity 2) is $\kappa_{2}$ which can be fabricated by pumping Er$^{3+}$ ions in microtoroid resonator of Er$^{3+}$-doped silica~\cite{Nat.Phys.10.394,Science.346.328,IEEEJQE.46.1626}. In the rotating frame defined by $V_{1}=\exp[-i\omega_{l}t(a_{1}^{\dagger}a_{1}+a_{2}^{\dagger}a_{2})]$, the transformed Hamiltonian $H_{1}=V_{1}^{\dagger}HV_{1}-iV_{1}^{\dagger}\dot{V_{1}}$ reads
\begin{eqnarray}\label{e02}
H_{1}&=&\Delta_{1}a_{1}^{\dagger}a_{1}+\Delta_{2}a_{2}^{\dagger}a_{2}
+\omega_{m}b^{\dagger}b+J(a_{1}^{\dagger}a_{2}+a_{1}a_{2}^{\dagger})\cr\cr
&&-ga_{1}^{\dagger}a_{1}(b^{\dagger}+b)+(Ea_{1}^{\dagger}+E^{\ast}a_{1}),
\end{eqnarray}
where $\Delta_{j}=\omega_{j}-\omega_{l}$ is the $j$-th cavity-laser detuning. In the displacement representation of the mechanical mode, defined by the canonical transformation $V_{2}=\exp[g/\omega_{m}a_{1}^{\dagger}a_{1}(b^{\dagger}-b)]$, the transformed Hamiltonian $H_{2}=V_{2}^{\dagger}H_{1}V_{2}$ is
\begin{eqnarray}\label{e03}
H_{2}&=&\Delta_{1}a_{1}^{\dagger}a_{1}+\Delta_{2}a_{2}^{\dagger}a_{2}
+\omega_{m}b^{\dagger}b-\frac{g^{2}}{\omega_{m}}(a_{1}^{\dagger}a_{1})^{2}\cr\cr
&&+J[a_{1}^{\dagger}a_{2}e^{-\frac{g}{\omega_{m}}(b^{\dagger}-b)}+a_{1}a_{2}^{\dagger}e^{\frac{g}{\omega_{m}}(b^{\dagger}-b)}]\cr\cr
&&+[Ea_{1}^{\dagger}e^{-\frac{g}{\omega_{m}}(b^{\dagger}-b)}+E^{\ast}a_{1}e^{\frac{g}{\omega_{m}}(b^{\dagger}-b)}].
\end{eqnarray}
Under the weak optomechanical coupling condition ($g\ll\omega_{m}$), those exponential factors of Eq.~(\ref{e03}) can be approximately omitted (all the approximates will be validated later via numerical simulation). Then the Hamiltonian can be rewritten as
\begin{eqnarray}\label{e04}
H_{3}&=&\Delta_{1}a_{1}^{\dagger}a_{1}+\Delta_{2}a_{2}^{\dagger}a_{2}+\omega_{m}b^{\dagger}b
-\frac{g^{2}}{\omega_{m}}(a_{1}^{\dagger}a_{1})^{2}\cr\cr
&&+J(a_{1}^{\dagger}a_{2}+a_{1}a_{2}^{\dagger})+(Ea_{1}^{\dagger}+E^{\ast}a_{1}),
\end{eqnarray}
which indicates that the mechanical resonator is decoupled from the system; namely, the evolutions of optical and mechanical parts are independent. So the state of the system is a separable state, which can be written as $|\psi\rangle|\hat{n}_{m}\rangle=\sum_{n_{1},n_{2}}C_{n_{1}n_{2}}|n_{1},n_{2}\rangle|\hat{n}_{m}\rangle$, where $n_{j}$ represents the photon number in $j$-th cavity and $|\hat{n}_{m}\rangle=V_{2}^{\dagger}|n_{m}\rangle$ is the mechanical displaced Fock state related to the photon number of cavity 1. In Hamiltonian (\ref{e04}), the part of the mechanical resonator can be ignored  when we only care about the optical properties of the system. The reduced Hamiltonian reads
\begin{eqnarray}\label{e05}
H_{4}&=&\Delta_{1}a_{1}^{\dagger}a_{1}+\Delta_{2}a_{2}^{\dagger}a_{2}
-\frac{g^{2}}{\omega_{m}}(a_{1}^{\dagger}a_{1})^{2}\cr\cr
&&+J(a_{1}^{\dagger}a_{2}+a_{1}a_{2}^{\dagger})+(Ea_{1}^{\dagger}+E^{\ast}a_{1}),
\end{eqnarray}
which is similar to the coupled cavity system consisting of a linear cavity and a Kerr-type nonlinear cavity~\cite{PhysRevA.80.065801,arXiv:1803.06642}, where the Kerr-type nonlinear strength is related to the mechanical resonator.

\section{Photon blockade in the $\mathcal{PT}$-symmetric optomechanical system}\label{sec.3}
\subsection{Analytical solution}\label{subsec.3A}
Here, the system we consider is a $\mathcal{PT}$-symmetric double-cavity optomechanical system, which includes a passive cavity and an active cavity. Naturally, the non-Hermitian Hamiltonian can be written by adding phenomenologically the imaginary decay and gain terms into the original Hamiltonian. Furthermore, due to the added imaginary terms have no effect on the previous calculation process, the reduced non-Hermitian Hamiltonian can be directly written as
\begin{eqnarray}\label{e06}
H_{\mathrm{NM}}&=&(\Delta_{1}-i\frac{\kappa_{1}}{2})a_{1}^{\dagger}a_{1}+(\Delta_{2}+i\frac{\kappa_{2}}{2})a_{2}^{\dagger}a_{2}
	-\frac{g^{2}}{\omega_{m}}(a_{1}^{\dagger}a_{1})^{2}\cr\cr
	&&+J(a_{1}^{\dagger}a_{2}+a_{1}a_{2}^{\dagger})+(Ea_{1}^{\dagger}+E^{\ast}a_{1}).
\end{eqnarray}

To satisfy the $\mathcal{PT}$-symmetry and simplify the next calculation, we set $\Delta_{1}=\Delta_{2}$ and $\kappa_{1}=\kappa_{2}$. The dynamical evolution of the optical components is calculated via the Schr\"{o}dinger equation $i\partial|\psi(t)\rangle/\partial t=H_{\mathrm{NM}}|\psi(t)\rangle$, where $|\psi(t)\rangle$ is the time-dependent photon state of the coupled cavities. Under the weak driving condition, we can restrict the total photon number within the low-excitation subspace up to 2. At this time, the time-dependent photon state $|\psi(t)\rangle$ can be expressed as
\begin{eqnarray}\label{e07}
|\psi(t)\rangle&=&C_{00}(t)|0,0\rangle+C_{10}(t)|1,0\rangle+C_{01}(t)|0,1\rangle\cr\cr
	&&+C_{20}(t)|2,0\rangle+C_{11}(t)|1,1\rangle+C_{02}(t)|0,2\rangle,
\end{eqnarray}
where $C_{n_{1}n_{2}}(t)$ is the probability amplitude for the coupled cavities being in the state $|n_{1},n_{2}\rangle$. Utilizing the above Schr\"{o}dinger equation, we can get a set of linear differential equations for the probability amplitudes
\begin{eqnarray}\label{e08}
i\frac{\partial C_{10}}{\partial t}&=&(\Delta_{1}-i\frac{\kappa_{1}}{2}-\frac{g^{2}}{\omega_{m}})C_{10}+JC_{01}+E+\sqrt{2}E^{\ast}C_{20},\cr\cr
i\frac{\partial C_{01}}{\partial t}&=&(\Delta_{1}+i\frac{\kappa_{1}}{2})C_{01}+JC_{10}+E^{\ast}C_{11},\cr\cr
i\frac{\partial C_{20}}{\partial t}&=&2(\Delta_{1}-i\frac{\kappa_{1}}{2}-\frac{2g^{2}}{\omega_{m}})C_{20}+\sqrt{2}JC_{11}+\sqrt{2}EC_{10},\cr\cr
i\frac{\partial C_{11}}{\partial t}&=&(2\Delta_{1}-\frac{g^{2}}{\omega_{m}})C_{11}+\sqrt{2}J(C_{20}+C_{20})+EC_{01},\cr\cr
i\frac{\partial C_{02}}{\partial t}&=&2(\Delta_{1}+i\frac{\kappa_{1}}{2})C_{02}+\sqrt{2}JC_{11}.
\end{eqnarray}
Here, it has the fact $\{C_{20},C_{11},C_{02}\}\ll\{C_{10},C_{01}\}\ll C_{00}$ and we thus can set $C_{00}\simeq1$ due to the weak driving assumption. Next, by ignoring higher-order terms $E^{\ast}C_{20}$ and $E^{\ast}C_{11}$, the steady-state solution can be obtained approximatively
\begin{eqnarray}\label{e09}
C_{10}&=&\frac{2\omega_{m}E(2\Delta_{1}+i\kappa_{1})}{2g^{2}(2\Delta_{1}+i\kappa_{1})-\omega_{m}(4\Delta_{1}^{2}+\kappa_{1}^{2}-4J^{2})},\cr\cr
C_{01}&=&\frac{-4J\omega_{m}E}{2g^{2}(2\Delta_{1}+i\kappa_{1})-\omega_{m}(4\Delta_{1}^{2}+\kappa_{1}^{2}-4J^{2})},\cr\cr
C_{20}&=&2\sqrt{2}\omega_{m}^{2}E^{2}(2\Delta_{1}+i\kappa_{1})^{2}(g^{2}-2\Delta_{1}\omega_{m})/M,\cr\cr
C_{11}&=&16J\omega_{m}^{2}E^{2}(2\Delta_{1}+i\kappa_{1})(\Delta_{1}\omega_{m}-g^{2})/M,\cr\cr
C_{02}&=&16\sqrt{2}J^{2}\omega_{m}^{2}E^{2}(g^{2}-\Delta_{1}\omega_{m})/M,
\end{eqnarray}
with
\begin{eqnarray}\label{e10}
M&=&[2g^{2}(2\Delta_{1}+i\kappa_{1})-\omega_{m}(4\Delta_{1}^{2}+\kappa_{1}^{2}-4J^{2})]\cr\cr
&&\times[4g^{4}(2\Delta_{1}+i\kappa_{1})+2\Delta_{1}\omega_{m}^{2}(4\Delta_{1}^{2}+\kappa_{1}^{2}-4J^{2})\cr\cr
&&-g^{2}\omega_{m}(20\Delta_{1}^{2}+8i\Delta_{1}\kappa_{1}+\kappa_{1}^{2}-8J^{2})].
\end{eqnarray}
Normally, the photon blockade effect is usually characterized by the equal-time second-order correlation function $g_{j}^{(2)}(0)=\langle a_{j}^{\dagger}a_{j}^{\dagger}a_{j}a_{j}\rangle/\langle a_{j}^{\dagger}a_{j}\rangle^{2}$, which characterizes the probability of detecting two photons at the same time. Here, we have the photon bunching effect for $g_{j}^{(2)}(0)>1$ and the photon antibunching effect for $g_{j}^{(2)}(0)<1$. On the other hand, the cross correlation function between the two cavities can be also calculated by $g_{12}^{(2)}(0)=\langle a_{1}^{\dagger}a_{2}^{\dagger}a_{2}a_{1}\rangle/(\langle a_{1}^{\dagger}a_{1}\rangle\langle a_{2}^{\dagger}a_{2}\rangle)$, which represents the probability that each cavity has one photon at the same time. Then the steady-state correlation functions of the two cavities can be analytically obtained via the steady-state solution of the Schr\"{o}dinger equation, respectively, and are given by
\begin{eqnarray}\label{e11}
g_{1}^{(2)}(0)&=&\frac{2|C_{20}|^{2}}{(|C_{10}|^{2}+|C_{11}|^{2}+2|C_{20}|^{2})^{2}}\simeq\frac{2|C_{20}|^{2}}{|C_{10}|^{4}},\cr\cr
g_{2}^{(2)}(0)&=&\frac{2|C_{02}|^{2}}{(|C_{01}|^{2}+|C_{11}|^{2}+2|C_{02}|^{2})^{2}}\simeq\frac{2|C_{02}|^{2}}{|C_{01}|^{4}},\cr\cr
g_{12}^{(2)}(0)&=&|C_{11}|^{2}/[(|C_{10}|^{2}+|C_{11}|^{2}+2|C_{20}|^{2})\cr\cr
				&&\times(|C_{01}|^{2}+|C_{11}|^{2}+2|C_{02}|^{2})]\cr\cr
			&\simeq&\frac{|C_{11}|^{2}}{|C_{10}|^{2}|C_{01}|^{2}}=\frac{2|C_{02}|^{2}}{|C_{01}|^{4}},
\end{eqnarray}
where $g_{1}^{(2)}(0)$ and $g_{2}^{(2)}(0)$ are the steady-state correlation functions of the passive cavity and the active cavity, respectively. $g_{12}^{(2)}(0)$ is the cross correlation function of the two cavities, which equals $g_{2}^{(2)}(0)$ approximately. Combining the above expressions with Eq.~(\ref{e09}), it is easy to find that the perfect photon blockade can be achieved in the $\mathcal{PT}$-symmetric double-cavity optomechanical system when the detuning $\Delta_{1}$ equals a determined value, which is related to the single photon optomechanical coupling strength and mechanical frequency. That is to say, when the detuning satisfies $\Delta_{1}=g^{2}/(2\omega_{m})$, the amplitude $C_{20}=0$, showing that the perfect photon blockade of the passive cavity 1 occurs. Similarly, when the detuning is set as $\Delta_{1}=g^{2}/\omega_{m}$, the correlation functions $g_{2}^{(2)}(0)=g_{12}^{(2)}(0)=0$ are obtained.

\subsection{Numerical simulation}
The exact solution of the photon blockade can also be obtained via numerical simulation, which utilizes the quantum master equation with the initial Hamiltonian $H_{1}$. The dynamics of the system is described by the quantum master equation as
\begin{eqnarray}\label{e12}
\dot{\rho}&=&-i[H_{1},\rho]+\kappa_{1}\mathcal{L}[a_{1}]\rho-\kappa_{2}\mathcal{L}[a_{2}]\rho\cr\cr
	&&+\gamma_{m}(n_\mathrm{th}+1)\mathcal{L}[b]\rho+\gamma_{m}n_\mathrm{th}\mathcal{L}[b^{\dagger}]\rho,
\end{eqnarray}
where $\mathcal{L}[o]\rho=o\rho o^{\dagger}-(o^{\dagger}o\rho+\rho o^{\dagger}o)/2$ is the standard Lindblad operator for the arbitrary system operator $o$. The negative sign in front of $\kappa_{2}$ indicates that the cavity 2 is an active cavity. $\gamma_{m}$ is the damping rate of the mechanical resonator. $n_\mathrm{th}=\{\exp[\hbar\omega_{m}/(k_{B}T)]-1\}^{-1}$ is the mean thermal excitation number of the mechanical resonator at temperature $T$, where $k_{B}$ is the Boltzmann constant. Under the limit of a long time, we can obtain the steady-state density matrix $\rho_{s}$ of the $\mathcal{PT}$-symmetric double-cavity optomechanical system. Meanwhile, the steady-state correlation functions are given by
\begin{eqnarray}\label{e13}
g_{1}^{(2)}(0)&=&\frac{\mathrm{Tr}(a_{1}^{\dagger}a_{1}^{\dagger}a_{1}a_{1}\rho_{s})}{[\mathrm{Tr}(a_{1}^{\dagger}a_{1}\rho_{s})]^{2}},\cr\cr
g_{2}^{(2)}(0)&=&\frac{\mathrm{Tr}(a_{2}^{\dagger}a_{2}^{\dagger}a_{2}a_{2}\rho_{s})}{[\mathrm{Tr}(a_{2}^{\dagger}a_{2}\rho_{s})]^{2}},\cr\cr
g_{12}^{(2)}(0)&=&\frac{\mathrm{Tr}(a_{1}^{\dagger}a_{2}^{\dagger}a_{2}a_{1}\rho_{s})}{\mathrm{Tr}(a_{1}^{\dagger}a_{1}\rho_{s})\mathrm{Tr}(a_{2}^{\dagger}a_{2}\rho_{s})}.
\end{eqnarray}

\begin{figure}
	\centering
	\includegraphics[width=0.49\linewidth]{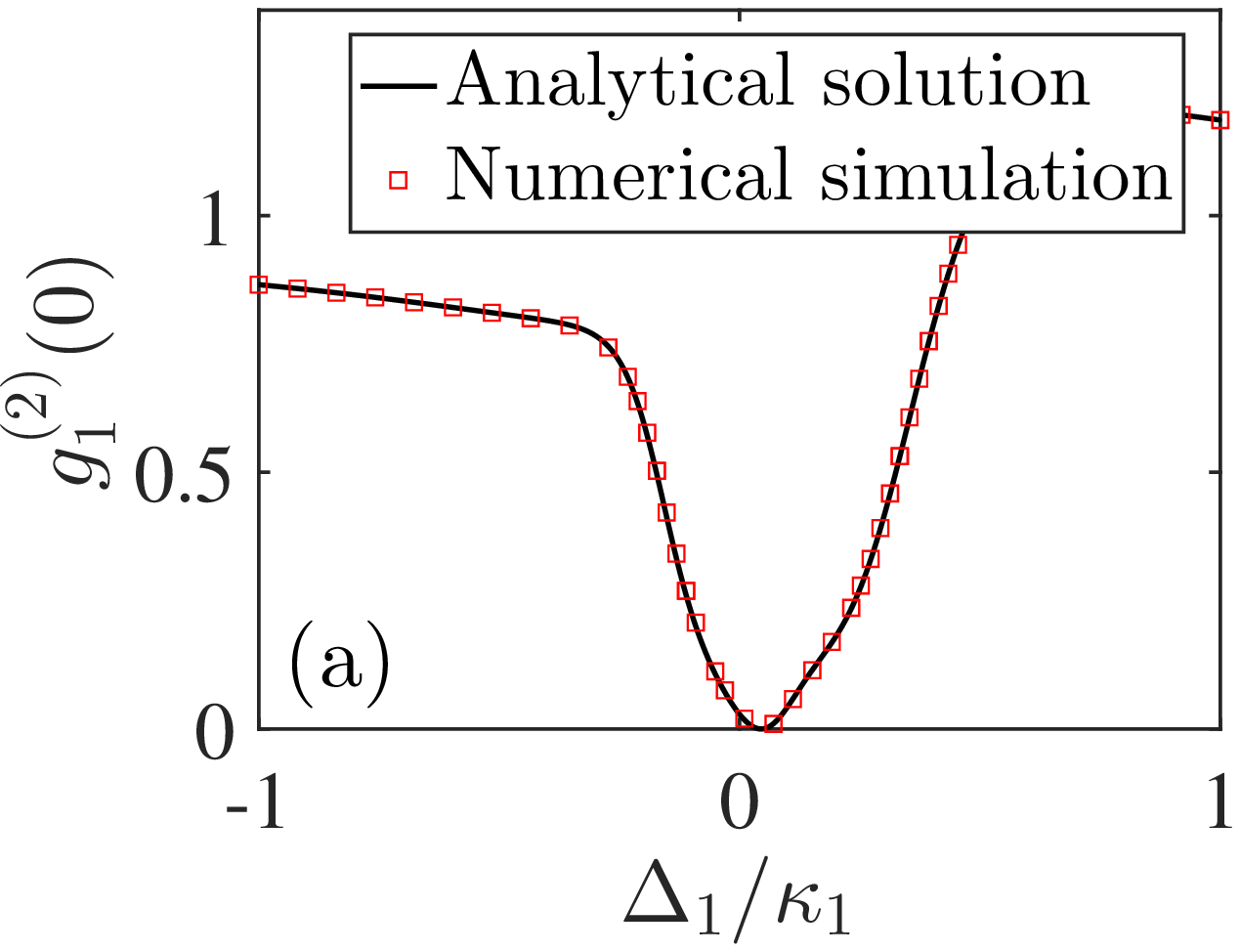}
	\hspace{0in}%
	\includegraphics[width=0.49\linewidth]{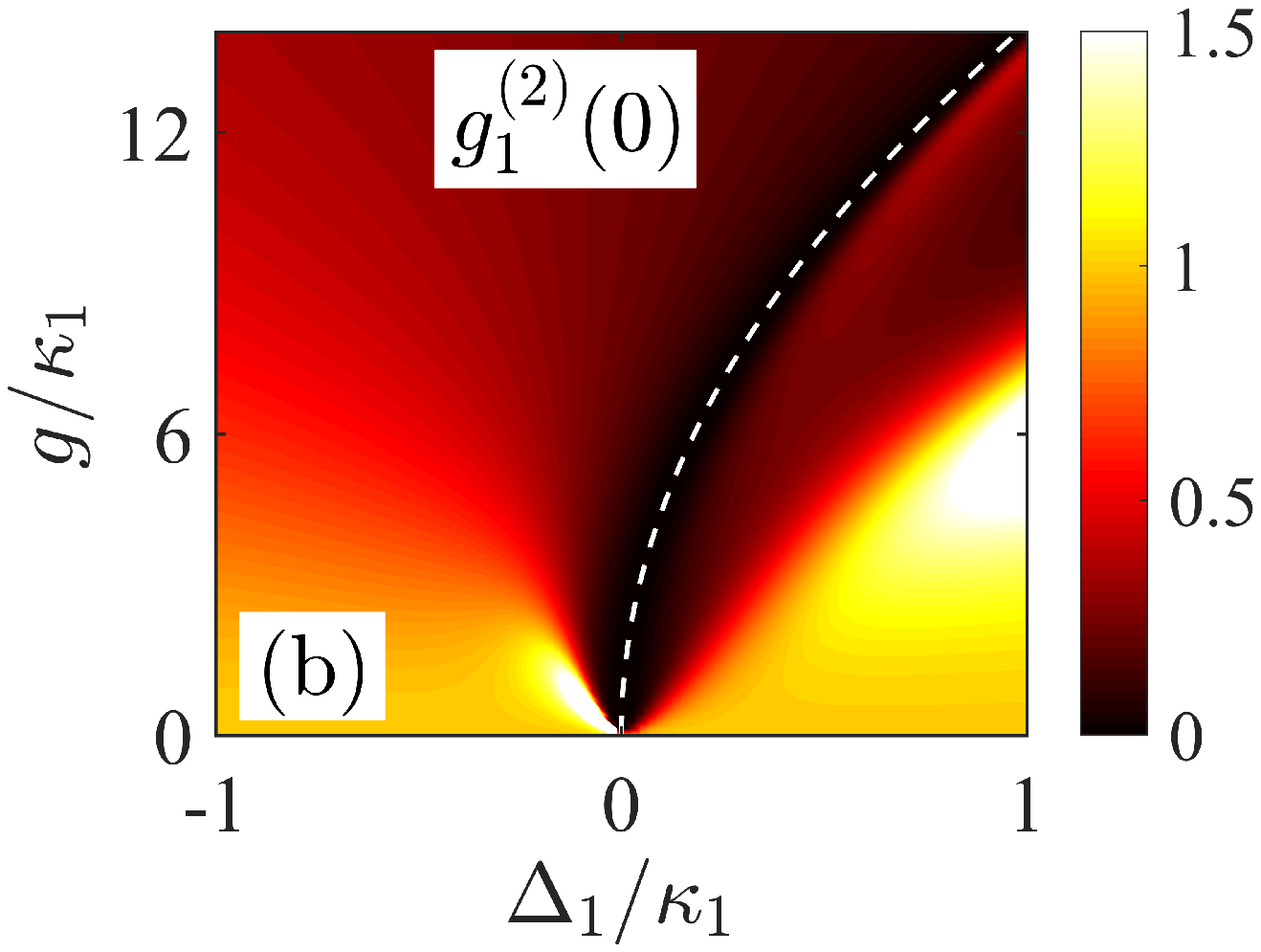}
	\vspace{0in}%
	\includegraphics[width=0.49\linewidth]{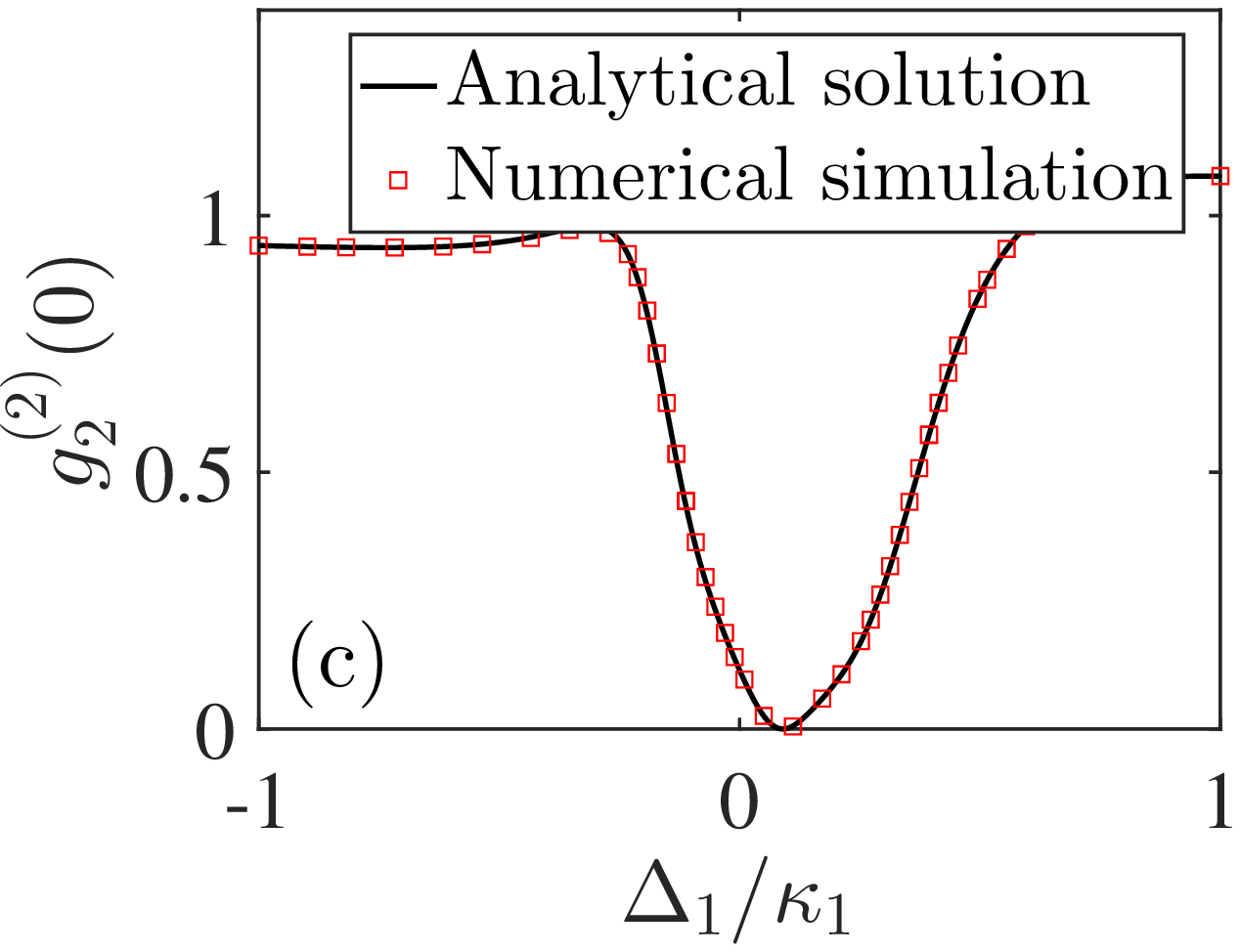}
	\hspace{0in}%
	\includegraphics[width=0.49\linewidth]{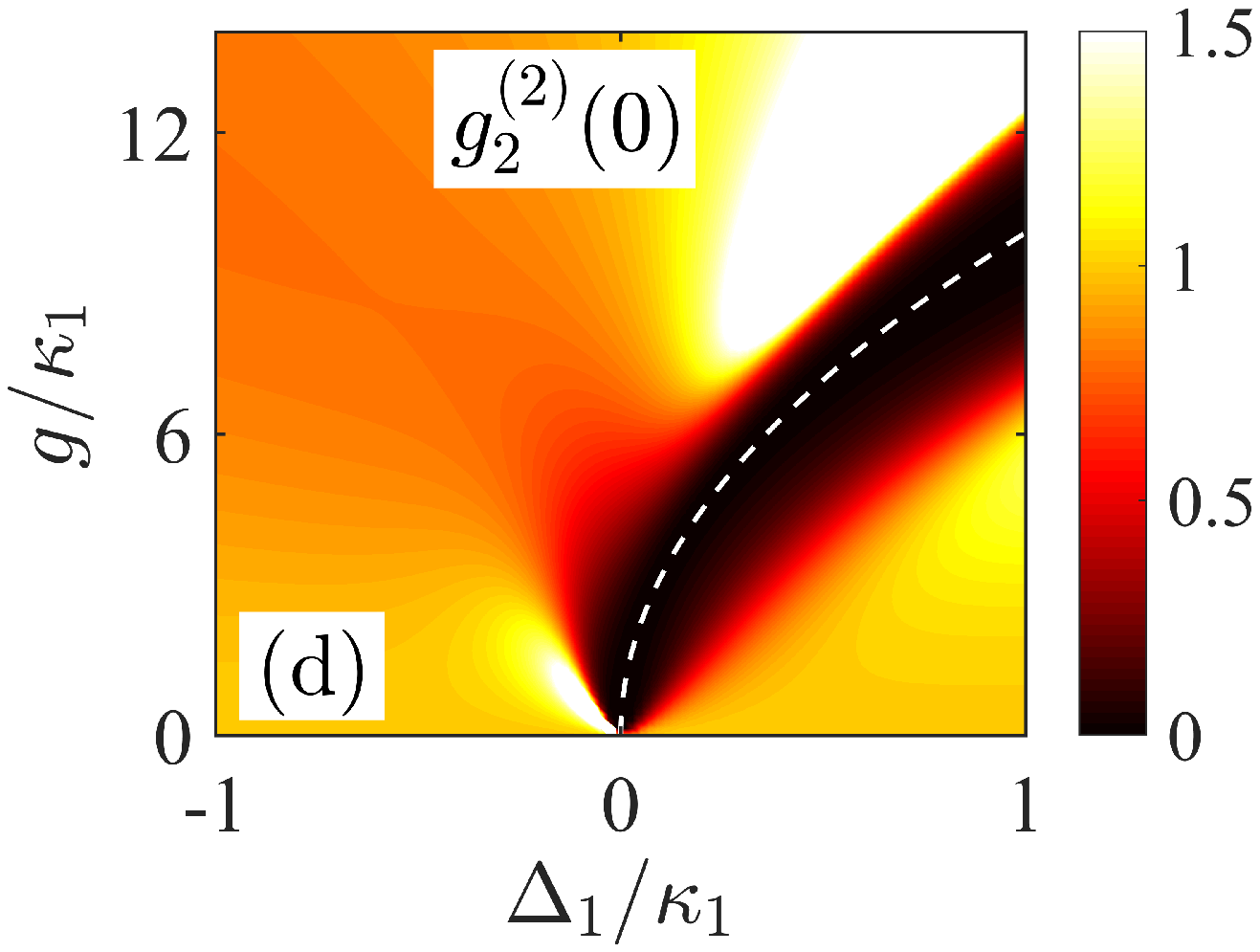}
	\vspace{0in}%
	\includegraphics[width=0.49\linewidth]{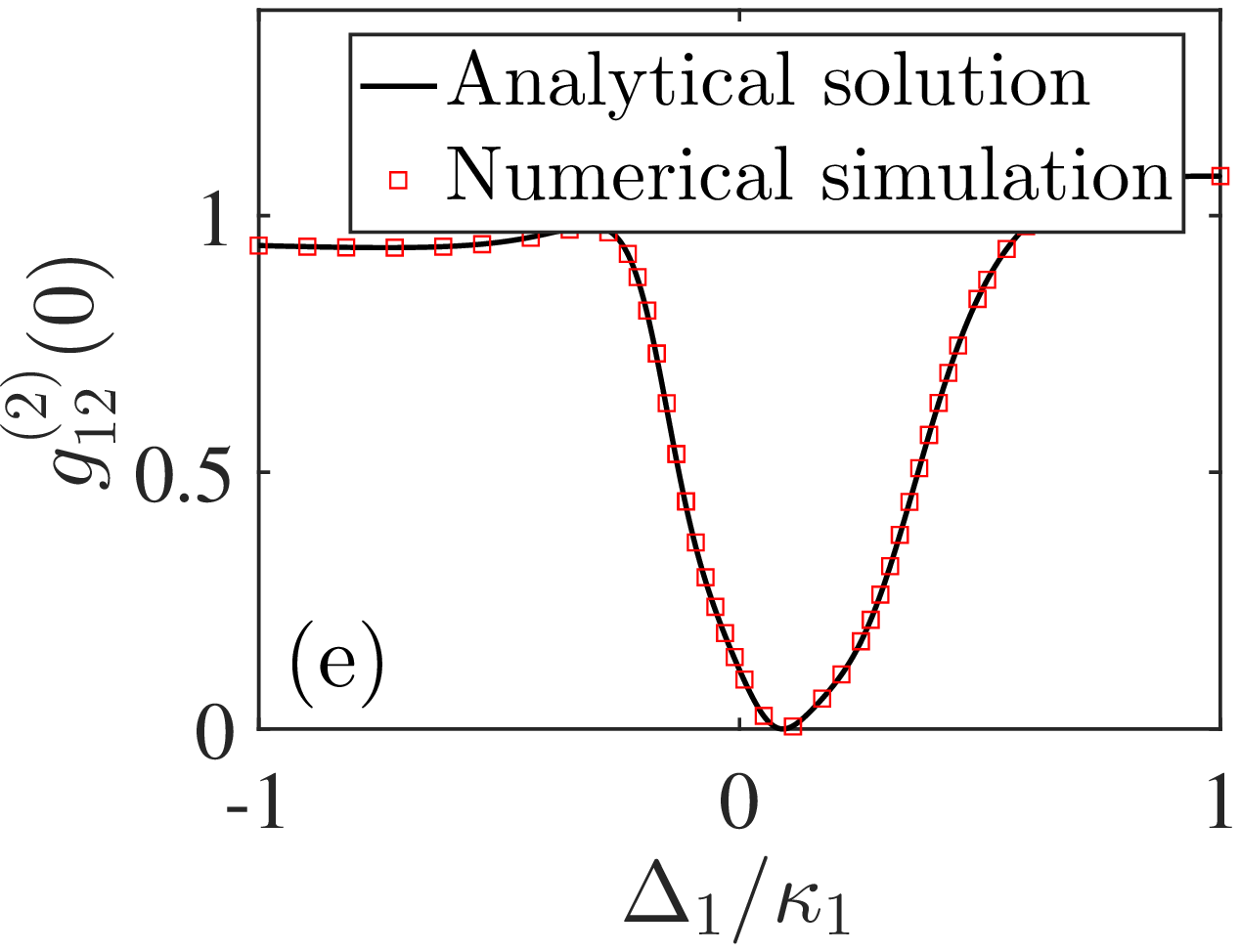}
	\hspace{0in}%
	\includegraphics[width=0.49\linewidth]{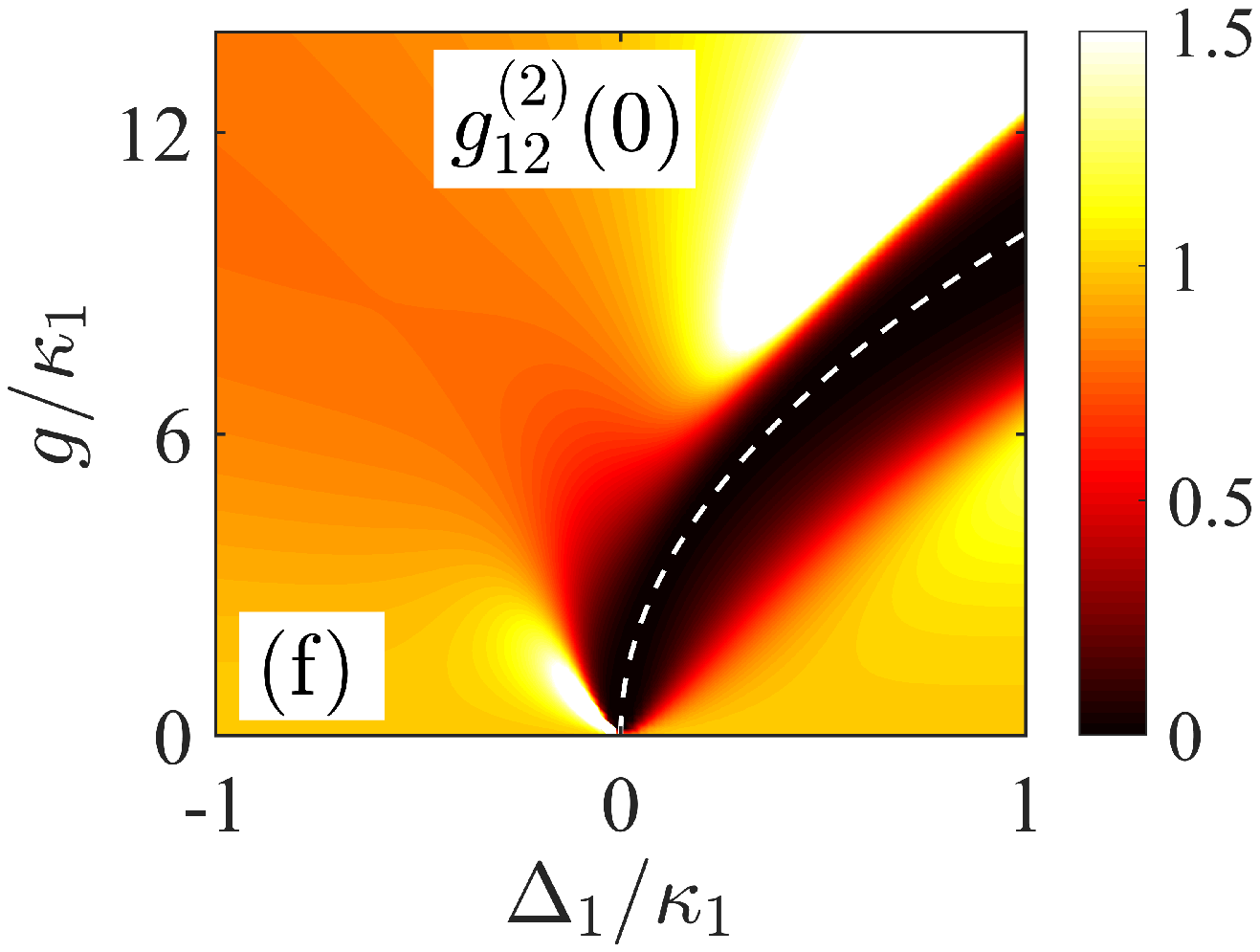}
	\caption{The equal-time second-order correlation functions versus the detuning (left column) or versus both the detuning and the optomechanical coupling strength (right column). In the left column, the solid black lines are the analytical solutions of the correlation functions defined as Eq.~(\ref{e11}) and the red squares represent the numerical simulation results by the master function Eq.~(\ref{e12}) with the Hamiltonian $H_{1}$, respectively. In the right column, the dashed white line is the optimal detuning changing with the optomechanical coupling strength. The system parameters are set as $T=1~\mathrm{mK}$, $\kappa_{1}=2\pi~\mathrm{MHz}$, $\kappa_{2}=\kappa_{1}$, $\Delta_{2}=\Delta_{1}$, $\omega_{m}=100\kappa_{1}$, $\gamma_{m}=\omega_{m}/10^{6}$, $J=0.5\kappa_{1}$, $g=3\kappa_{1}$, and $E=0.01\kappa_{1}$.}
	\label{fig:blockade-Delta-g}
\end{figure}

\begin{figure}
	\centering
	\includegraphics[width=0.5\linewidth]{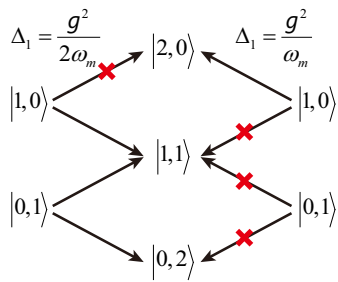}
	\caption{The perfect photon blockade with the different detunings.}
	\label{fig:Blockade}
\end{figure}

Next, we strictly validate the validity of our previous calculations by comparing all the results of correlation functions, which come from the analytical solutions and the numerical simulations with the Hamiltonian $H_{1}$ under the weak driving condition (see the left column in Fig.~\ref{fig:blockade-Delta-g}). The numerical simulations are in good agreement with the analytical solutions for the correlation functions $g_{1}^{(2)}(0)$, $g_{2}^{(2)}(0)$, and $g_{12}^{(2)}(0)$. In addition, the correlation functions versus both the detuning and the optomechanical coupling strength are shown in the right column of Fig.~\ref{fig:blockade-Delta-g}. The results indicate that the perfect photon blockade can be achieved when the detuning satisfies the optimal condition: $\Delta_{1}=g^{2}/(2\omega_{m})$ for $g_{1}^{(2)}(0)$ and $\Delta_{1}=g^{2}/\omega_{m}$ for $g_{2}^{(2)}(0)$ and $g_{12}^{(2)}(0)$ (see the dashed white line in the right column of Fig.~\ref{fig:blockade-Delta-g}). It is worth noting that the perfect photon blockade can be achieved with the weak single photon optomechanical coupling strength ($g\ll\omega_{m}$), which breaks the limits of strong coupling in the usual single-cavity optomechanical systems~\cite{PhysRevLett.107.063601,PhysRevLett.107.063602,PhysRevA.87.025803,PhysRevA.88.023853,PhysRevA.92.033806,PhysRevA.93.063860,arXiv:1802.09254}. Moreover, there has an interesting phenomenon due to the similar behaviors of the correlation functions $g_{2}^{(2)}(0)$ and $g_{12}^{(2)}(0)$, i.e., $g_{2}^{(2)}(0)=g_{12}^{(2)}(0)=0$ when the detuning satisfies $\Delta_{1}=g^{2}/\omega_{m}$. The excitations of both cavities are blocked simultaneously when the state of the coupled cavities is $|0,1\rangle$. That means the system no longer absorbs energy to excite the optical cavities until the state of the coupled cavities is changed. That is to say, when the only photon is transmitted to cavity 1, the system might be possible to be further excited through the cavity 1. On the other hand, for $\Delta_{1}=g^{2}/(2\omega_{m})$, the perfect blockade occurs only in the cavity 1. This means that when the state of the coupled cavities is $|1,0\rangle$, the system can be further excited only through the cavity 2. However, when the state of the coupled cavities is $|0,1\rangle$, the system can be further excited through any one of the two cavities. The phenomenon is clearly shown in Fig.~\ref{fig:Blockade}. It is worth noting that the imperfect photon blockade of any cavity can still be obtained in the vicinity of the optimal detuning due to the fact that all the correlation functions are less than 1.

\begin{figure}
	\centering
	\includegraphics[width=0.9\linewidth]{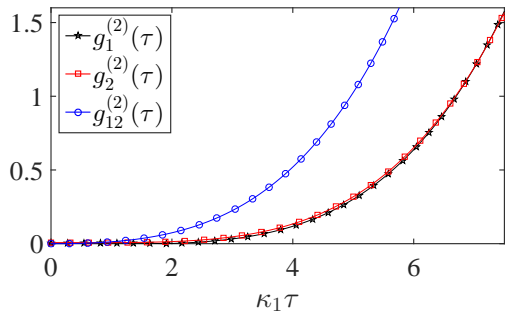}
	\caption{The delayed second-order correlation functions for different cavities versus the time delay $\kappa_{1}\tau$, where the detuning is respectively set as $\Delta_{1}=g^{2}/(2\omega_{m})$ for $g_{1}^{(2)}(\tau)$ and $\Delta_{1}=g^{2}/\omega_{m}$ for $g_{2}^{(2)}(\tau)$ and $g_{12}^{(2)}(\tau)$. The other parameters are same as in Fig.~\ref{fig:blockade-Delta-g}.}
	\label{fig:twotime}
\end{figure}

Finally, to characterize the nature of the photon emission, we also calculate the delayed second-order correlation functions in the steady state. The delayed second-order correlation functions are defined by $g_{j}^{(2)}(\tau)=\langle a_{j}^{\dagger}(0)a_{j}^{\dagger}(\tau)a_{j}(\tau)a_{j}(0)\rangle/\langle a_{j}^{\dagger}(0)a_{j}(0)\rangle^{2}$, where $\tau$ is a finite-time delay. The results of the delayed second-order correlation function for different cavities are shown in Fig.~\ref{fig:twotime}. One can see from Fig.~\ref{fig:twotime} that the value of the delayed second-order correlation function gradually increases with the increase of the time delay and it always exceeds 1 when the time delay is long enough. The main reason for this is the existence of the gain in $\mathcal{PT}$-symmetric double-cavity optomechanical system. While for the system consisting only decay cavities without gain, the delayed second-order correlation function of the no-gain system is finally stabilized at 1 with the increase of the time delay~\cite{PhysRevA.90.023849,PhysRevA.92.023838,PhysRevA.96.053810}. The deeper physical mechanism of the present results needs to be further studied and explored in the future.

\subsection{The effect of $\mathcal{PT}$ phase transition}
As we all know, the $\mathcal{PT}$-symmetric systems exist the interesting spontaneous $\mathcal{PT}$-symmetry breaking behaviour, which comes from the variation of the system parameters and it is determined via checking the eigenvalue. In order to study the effect of $\mathcal{PT}$ phase transition on the photon blockade, we should first find the threshold of the spontaneous $\mathcal{PT}$-symmetry breaking. To this end, we expand the non-Hermitian Hamiltonian without the driving terms, $H_{\mathrm{ND}}=(\Delta_{1}-i\kappa_{1}/2)a_{1}^{\dagger}a_{1}+(\Delta_{2}+i\kappa_{2}/2)a_{2}^{\dagger}a_{2}
-g^{2}/\omega_{m}(a_{1}^{\dagger}a_{1})^{2}+J(a_{1}^{\dagger}a_{2}+a_{1}a_{2}^{\dagger})$, by the vector $a=[a_{1},a_{2}]^{T}$. Ignoring the weak nonlinear term $g^{2}/\omega_{m}(a_{1}^{\dagger}a_{1})^{2}$, the result reads
\begin{eqnarray}\label{e14}
h=\left(\begin{array}{cc}
\Delta_{1}-i\frac{\kappa_{1}}{2}~&~J\\ 
J & \Delta_{2}+i\frac{\kappa_{2}}{2}
\end{array}\right),
\end{eqnarray}
where $a^{\dagger}ha$ is the approximate Hamiltonian. Under the condition of satisfying $\mathcal{PT}$-symmetry, i.e., $\Delta_{1}=\Delta_{2}$ and $\kappa_{1}=\kappa_{2}$, the eigenvalues of the matrix $h$ are $\varepsilon_{\pm}=\Delta_{1}\pm\sqrt{J^{2}-(\kappa_{1}/2)^{2}}$, which are real only when $J\geqslant\kappa_{1}/2$. So the $\mathcal{PT}$ phase transition point is approximately $J=\kappa_{1}/2$, which is also called as exceptional point. Meanwhile, the $\mathcal{PT}$ symmetry can be roughly divided into two different regions: the unbroken ($J>\kappa_{1}/2$) and broken ($J<\kappa_{1}/2$) $\mathcal{PT}$-symmetric regions.

\begin{figure}
	\centering
	\includegraphics[width=0.49\linewidth]{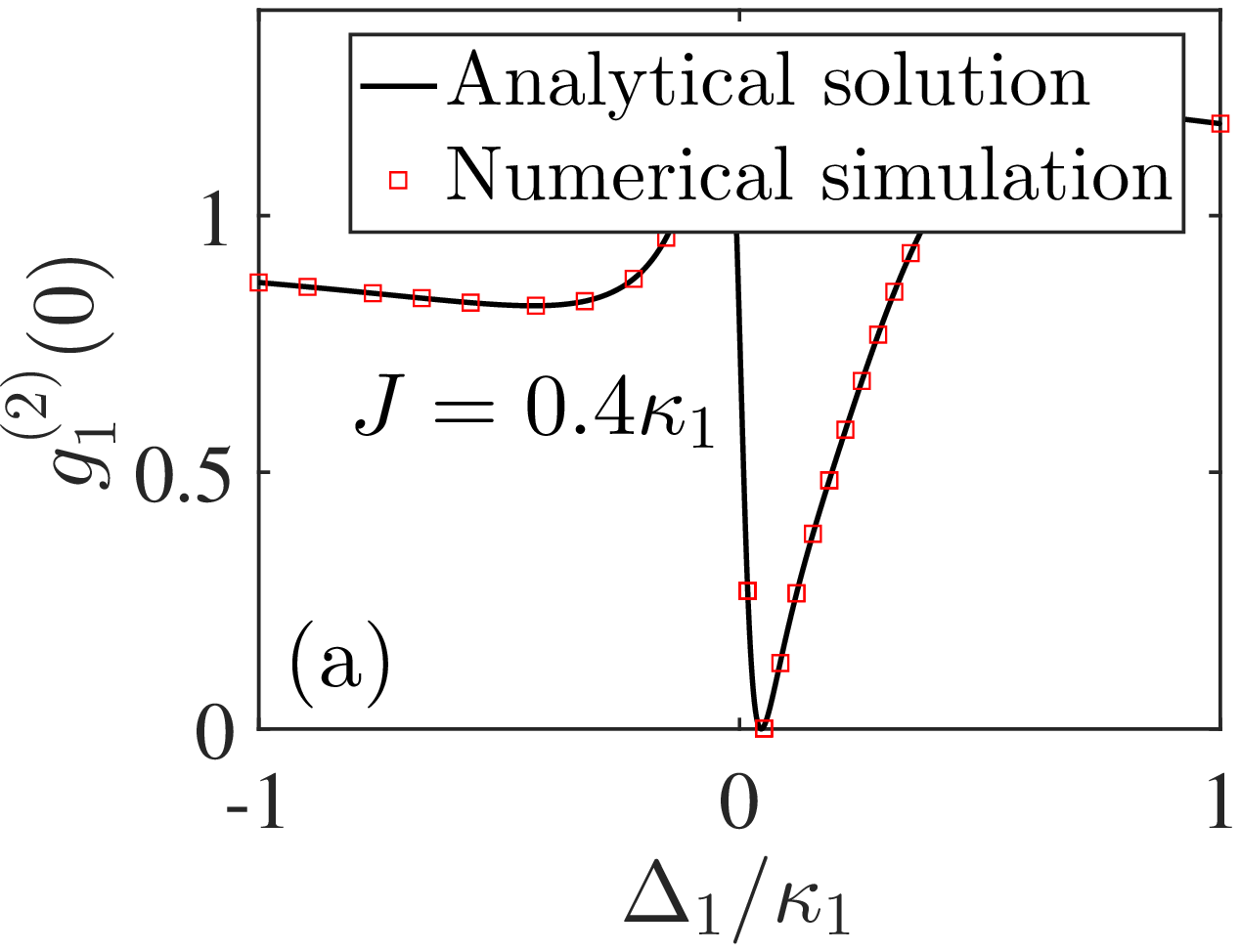}
	\hspace{0in}%
	\includegraphics[width=0.49\linewidth]{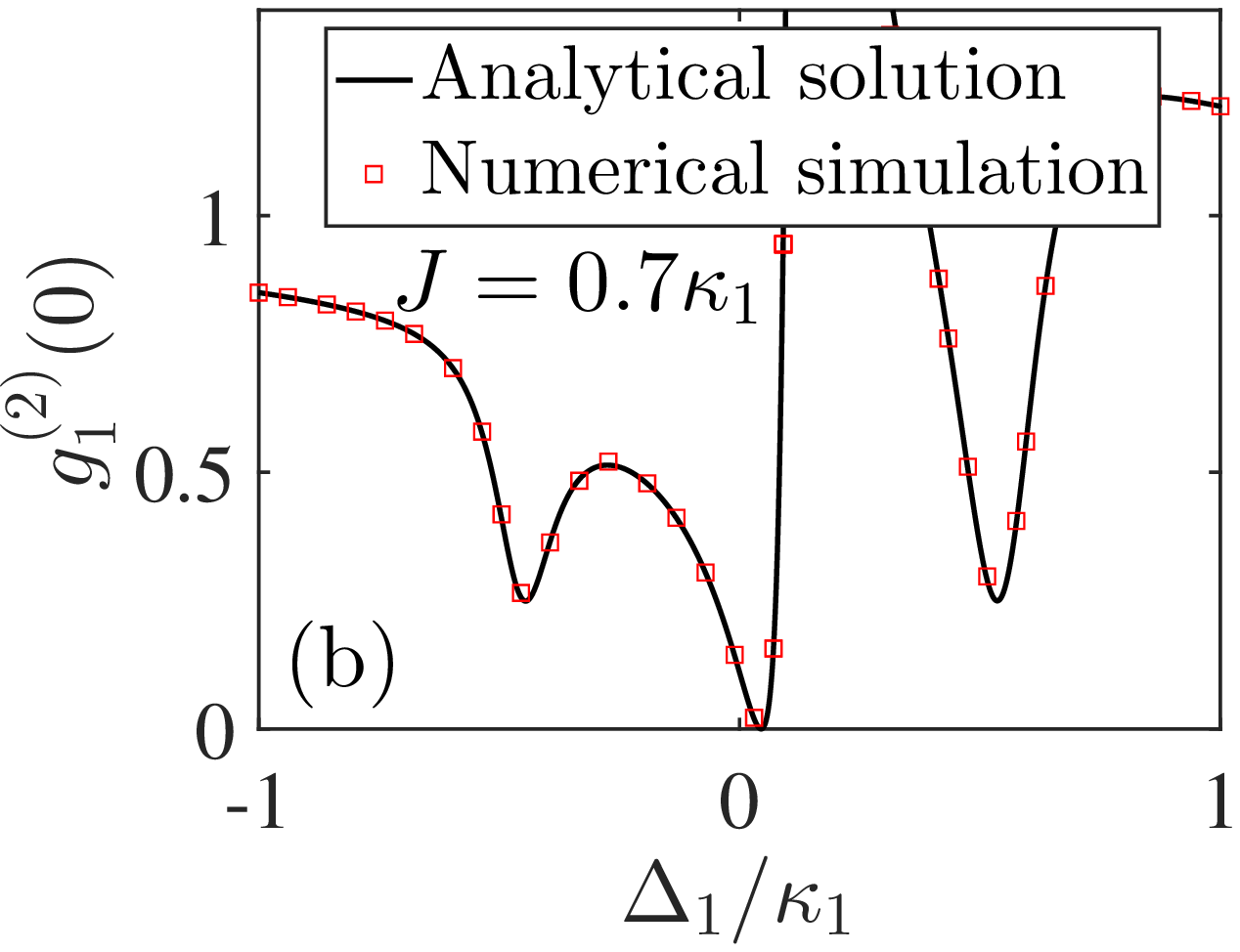}
	\vspace{0in}%
	\includegraphics[width=0.49\linewidth]{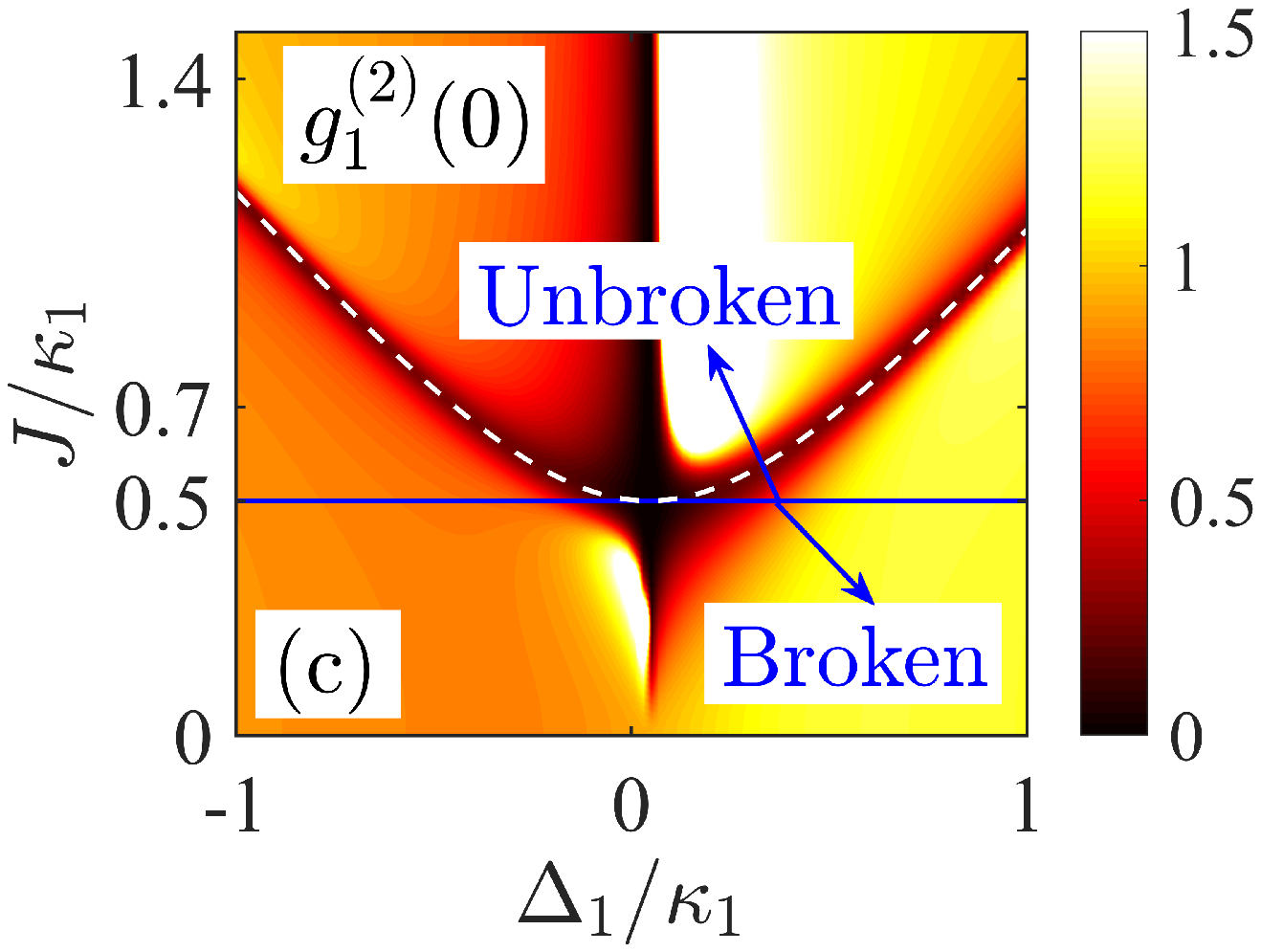}
	\hspace{0in}%
	\includegraphics[width=0.49\linewidth]{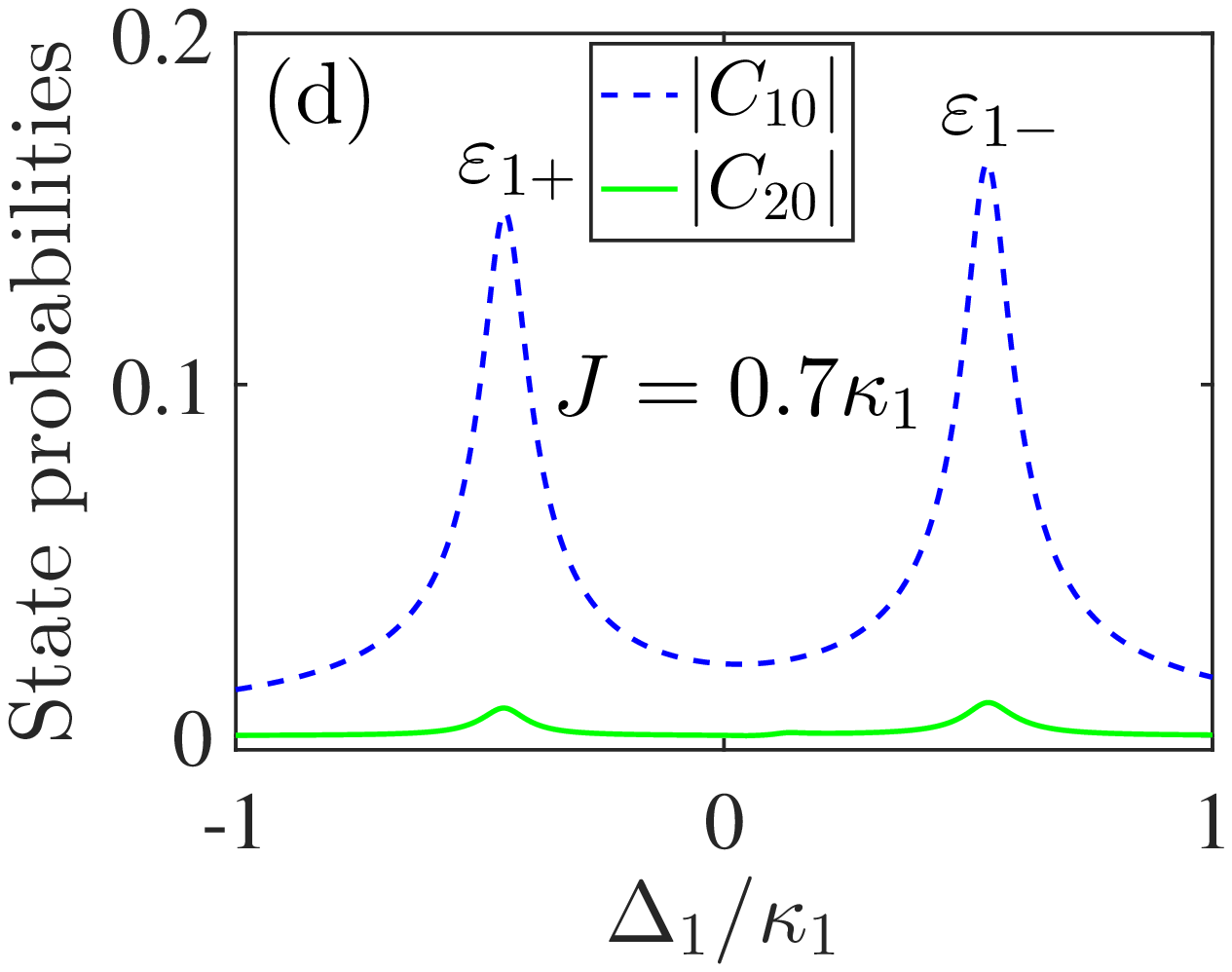}
	\caption{The equal-time correlation function of the passive cavity 1 in the different $\mathcal{PT}$-symmetric phases. (a) and (b) represent the correlation function changing with the detuning for different $J$, which correspond to the broken $\mathcal{PT}$-symmetric region and the unbroken $\mathcal{PT}$-symmetric region, respectively. The solid black line is the analytical solution of the correlation function $g_{1}^{(2)}(0)$ defined in Eq.~(\ref{e11}), and the red squares represent the numerical simulation result by the master function Eq.~(\ref{e12}) with the Hamiltonian $H_{1}$. (c) The correlation function $g_{1}^{(2)}(0)$ versus both the detuning and the coupling strength between the two cavities. The dashed white line is the optimal parameters, which come form the theory of CPB. The solid blue line represents the $\mathcal{PT}$ phase transition, which divides the system into the broken and unbroken $\mathcal{PT}$-symmetric regions. (d) The distributed probabilities of states $|1,0\rangle$ and $|2,0\rangle$ versus the detuning for $J=0.7\kappa_{1}$. The other parameters are same as in Fig.~\ref{fig:blockade-Delta-g}.}
	\label{fig:blockade-Delta-J}
\end{figure}

In Fig.~\ref{fig:blockade-Delta-J}, we show the equal-time second-order correlation function of the passive cavity 1 in the different $\mathcal{PT}$-symmetric regions, where Fig.~\ref{fig:blockade-Delta-J}(a) belongs to the broken $\mathcal{PT}$-symmetric region and Fig.~\ref{fig:blockade-Delta-J}(b) belongs to the unbroken $\mathcal{PT}$-symmetric region. It is easy to find that the photon blockade behaviors are different in the different $\mathcal{PT}$-symmetric regions. One can see from Fig.~\ref{fig:blockade-Delta-J}(a) that there is one dip in the broken $\mathcal{PT}$-symmetric region, while three dips occur in the unbroken $\mathcal{PT}$-symmetric region, as shown in Fig.~\ref{fig:blockade-Delta-J}(b). To clearly show the effect of the $\mathcal{PT}$ phase transition on the correlation function of the passive cavity 1, we also show the correlation function changing with the detuning and the photon tunneling coupling strength in Fig.~\ref{fig:blockade-Delta-J}(c). It is clear to find that, except for the dip located at the optimal detuning (the middle one), the other two dips on both sides of the optimal detuning occur only when the system is in the unbroken $\mathcal{PT}$-symmetric region.

To explain the photon blockade effect in the $\mathcal{PT}$-symmetric double-cavity optomechanical system, we expand the non-Hermitian Hamiltonian without the driving term in different excitation subspaces. In the zero-excitation subspace, the eigenvalue equation of the system is $H_{\mathrm{ND}}|\phi_{0}\rangle=\varepsilon_{0}|\phi_{0}\rangle$, where $|\phi_{0}\rangle$ is the eigenstate and $\varepsilon_{0}=0$ is the eigenvalue. Similarly, in the single-excitation subspace $\{|1,0\rangle,|0,1\rangle\}$, the eigenvalue equation of the system is $H_{\mathrm{ND}}|\phi_{1\pm}\rangle=\varepsilon_{1\pm}|\phi_{1\pm}\rangle$ and the matrix form of $H_{\mathrm{ND}}$ is written as
\begin{eqnarray}\label{e15}
H_{\mathrm{ND}}=\left(\begin{array}{cc}
\Delta_{1}-i\frac{\kappa_{1}}{2}-\frac{g^{2}}{\omega_{m}}~&~J\\ 
J & \Delta_{2}+i\frac{\kappa_{2}}{2}
\end{array}\right).
\end{eqnarray}
Under the condition of satisfying $\mathcal{PT}$-symmetry, i.e., $\Delta_{1}=\Delta_{2}$ and $\kappa_{1}=\kappa_{2}$, the eigenvalues $\varepsilon_{1\pm}$ can be given approximatively as
\begin{eqnarray}\label{e16}
\varepsilon_{1\pm}\simeq\Delta_{1}-\frac{g^{2}}{2\omega_{m}}\pm\sqrt{J^2-(\frac{\kappa_{1}}{2})^{2}}.
\end{eqnarray}

\begin{figure}
	\centering
	\includegraphics[width=0.8\linewidth]{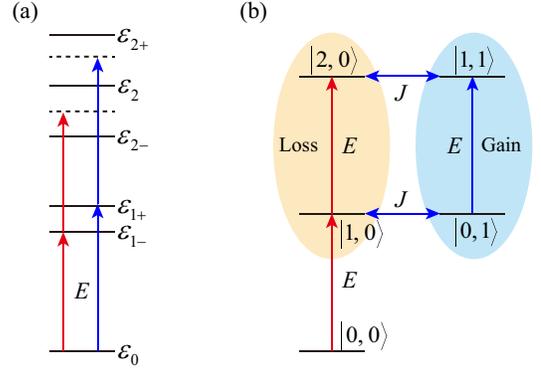}
	\caption{Level diagram of the photon state for the $\mathcal{PT}$-symmetric optomechanical system. (a) Dressed state. (b) Bare state.}
	\label{fig:level}
\end{figure}

According to the theory of CPB, for the resonant transitions between the zero-excitation and single-excitation states [see Fig.~\ref{fig:level}(a)], we can obtain the optimal relations $\varepsilon_{1\pm}-\varepsilon_{0}=0$ for the photon blockade. Moreover, the photon transitions of the higher-excitation state is off-resonant due to the anharmonicity of the eigenenergy spectrum. So the optimal photon blockade occurs when the system parameters satisfy the optimal relations. This can be verified via showing the probability amplitude of the single-excitation state, as shown in Fig.~\ref{fig:blockade-Delta-J}(d). We can see that, when the single-excitation is resonant, the amplitude of the state $|1,0\rangle$ reaches the peaks, which correspond to the resonance of different single-excitation eigenvalues, respectively. For the same photon tunneling coupling $J$, the locations of the peaks for the amplitude of the single-excitation state correspond to the locations of the optimal CPB [see Fig.~\ref{fig:blockade-Delta-J}(b,d)]. The optimal relations for the single-excitation resonance are also shown in Fig.~\ref{fig:blockade-Delta-J}(c) (see the dashed white line) in which they agree very well with the results of the photon blockade. Therefore, we can conclude that the two dips of the correlation function located at both side come from the single-excitation resonance, and belong to the CPB and only occur in the unbroken $\mathcal{PT}$-symmetric region.

\begin{figure}
	\centering
	\includegraphics[width=0.9\linewidth]{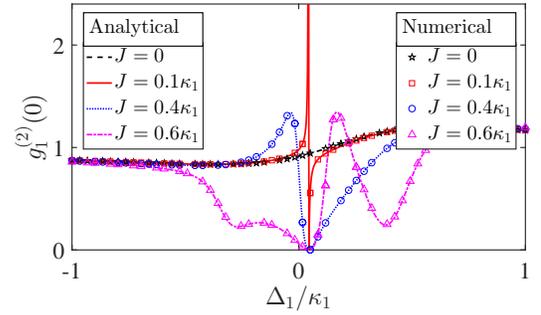}
	\caption{The equal-time correlation function of the passive cavity 1 versus the detuning with different $J$. The other parameters are the same as in Fig.~\ref{fig:blockade-Delta-g}.}
	\label{fig:ga1-Delta-diff-J}
\end{figure}

On the other hand, the middle dip of the correlation function $g_{1}^{(2)}(0)$ is located at the optimal detuning $\Delta_{1}=g^{2}/(2\omega_{m})$ whether the $\mathcal{PT}$-symmetry is broken or not, as shown in Fig.~\ref{fig:blockade-Delta-J}(a-c). To explore the reason for the photon blockade located at the optimal detuning, we draw the energy level of the bare photon state in Fig.~\ref{fig:level}(b). According to the theory of UPB, for the destructive interference between different paths of two-photon excitation, we show the different paths with different color arrows. For the $\mathcal{PT}$-symmetric double-cavity optomechanical system, the two paths suffer from different effects according to the $\mathcal{PT}$-symmetry, i.e., photon loss in the path $|1,0\rangle\rightarrow|2,0\rangle$ and photon gain in the path $|1,0\rangle\rightarrow|0,1\rangle\rightarrow|1,1\rangle\rightarrow|2,0\rangle$. That is different from the usual UPB. Through the calculation in Sec.~\ref{subsec.3A}, we find that the perfect photon blockade of arbitrary cavity can be achieved when the detuning is chosen properly. Although it is slightly different from the usual UPB, the perfect photon blockade located at the optimal detuning also come from the destructive interference between different paths, where the photon gain enhances the interference path resulting in the perfect photon blockade occurring even with the weak parameter mechanism. That can be further found in Fig.~\ref{fig:ga1-Delta-diff-J}, which show that the correlation function of the passive cavity 1 $g_{1}^{(2)}(0)$ changes with the detuning for different photon tunneling coupling strengths ($J/\kappa_{1}=0,~0.1,~0.4,~0.6$). We can see that, when the system is reduced to single-cavity optomechanical system ($J/\kappa_{1}=0$), the perfect photon blockade is non-existent (see the dashed black line and black pentagram line in Fig.~\ref{fig:ga1-Delta-diff-J}). But when the second active cavity is added, the perfect photon blockade occurs at the optimal detuning. And the width of dip located at optimal detuning increases with the increase of the photon tunneling coupling strength. When the coupling strength continues to increase to be larger than the $\mathcal{PT}$ phase transition point, the other two dips of the correlation function occur just as in the unbroken $\mathcal{PT}$-symmetric region (see the pecked magenta line and magenta triangle in Fig.~\ref{fig:ga1-Delta-diff-J}). In contrast to the usual UPB, the present photon blockade is slightly different, where the interference paths are affected by $\mathcal{PT}$-symmetry and the perfect photon blockade can be achieved even with the weak parameter mechanism.

\section{Non-$\mathcal{PT}$ symmetry}\label{sec.4}
In the above section, we study the photon blockade phenomenon in $\mathcal{PT}$-symmetric double-cavity optomechanical system and find the different blockade behaviors when the $\mathcal{PT}$ phase transition occurs. Here, as a comparison, we focus on the study of the photon blockade in non-$\mathcal{PT}$ symmetric double-cavity optomechanical system. There are two kinds of non-$\mathcal{PT}$ symmetric situations: $\kappa_{1}\neq\kappa_{2}$ or $\Delta_{1}\neq\Delta_{2}$.

\begin{figure}
	\centering
	\includegraphics[width=0.49\linewidth]{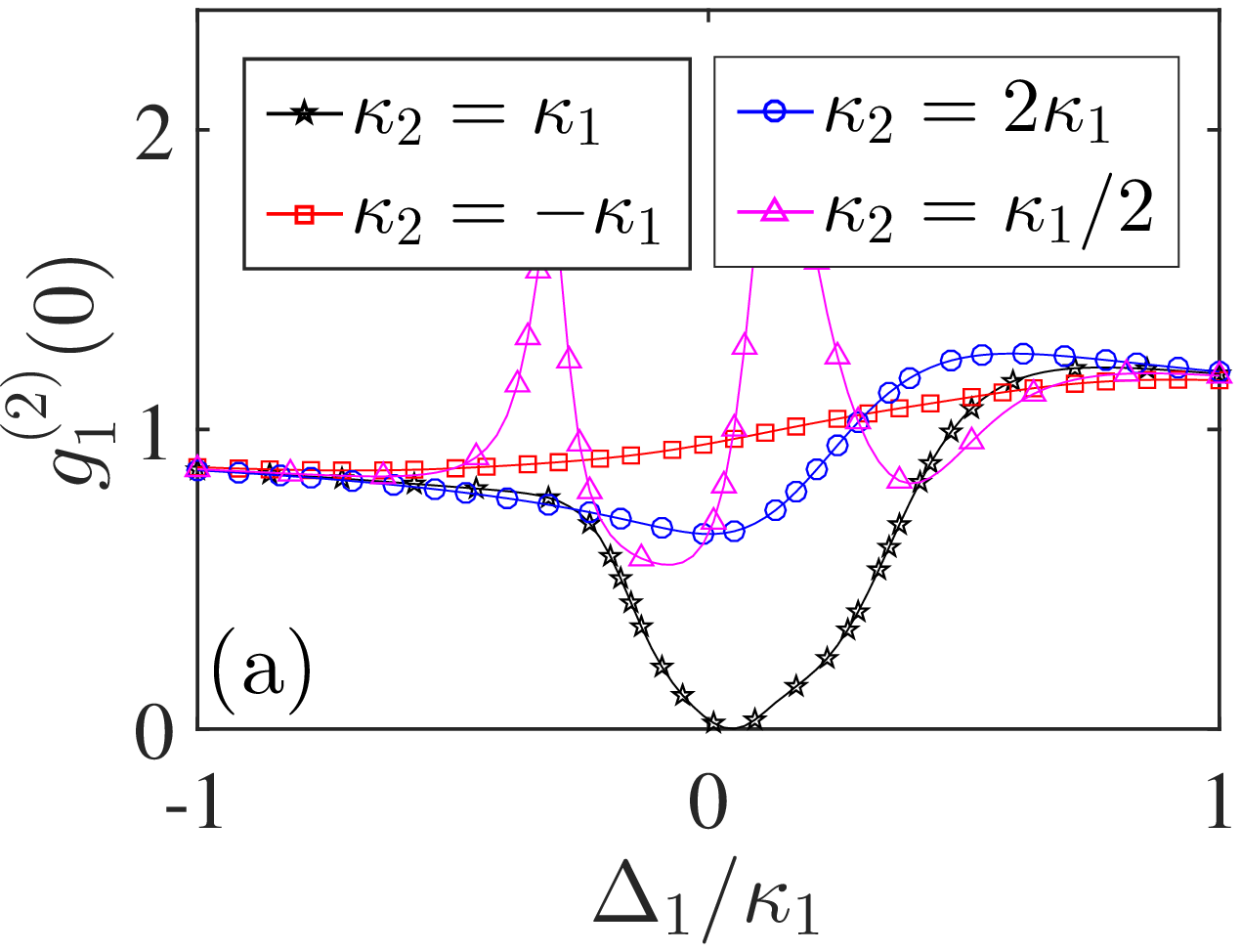}
	\hspace{0in}%
	\includegraphics[width=0.49\linewidth]{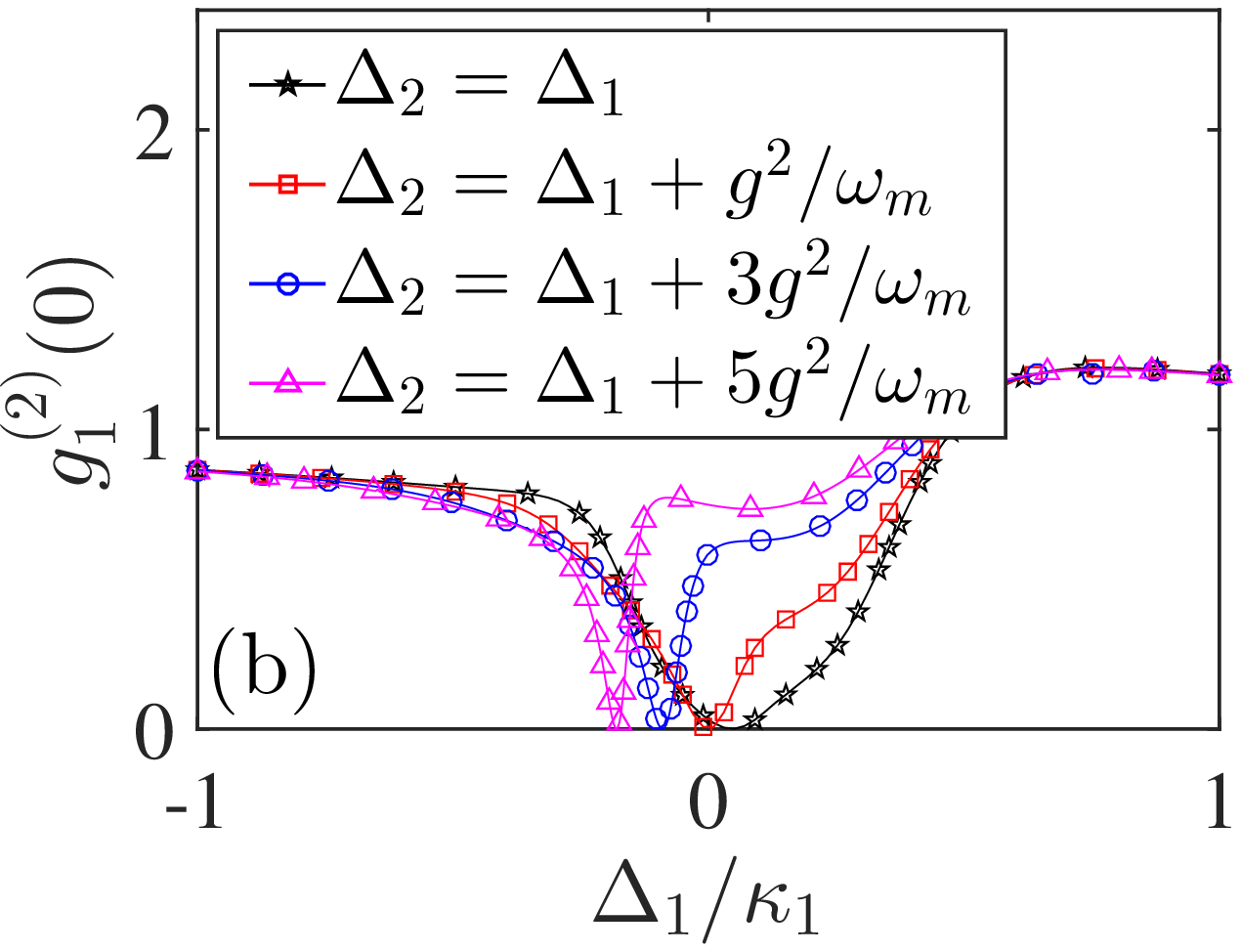}
	\caption{The equal-time correlation function of the passive cavity 1 versus the detuning in non-$\mathcal{PT}$ symmetric optomechanical system. (a) $\kappa_{1}\neq\kappa_{2}$ and $\Delta_{1}=\Delta_{2}$. (b) $\kappa_{1}=\kappa_{2}$ and $\Delta_{1}\neq\Delta_{2}$. The other parameters are the same as in Fig.~\ref{fig:blockade-Delta-g}.}
	\label{fig:ga1-Delta1-non-PT}
\end{figure}

For the first kind of the non-$\mathcal{PT}$ symmetric double-cavity optomechanical system $\kappa_{1}\neq\kappa_{2}$, we calculate the correlation function of the cavity 1 via simulating the master equation [Eq.~(\ref{e12})] numerically, and show the results in Fig~\ref{fig:ga1-Delta1-non-PT}(a). The results indicate that the perfect photon blockade located at the optimal detuning is achieved only when the gain-loss ratio is balance [see the black pentagram line in Fig.~\ref{fig:ga1-Delta1-non-PT}(a)]. However, when both of the cavities are passive, the photon blockade phenomenon in the cavity 1 is unobvious [see the red square line in Fig.~\ref{fig:ga1-Delta1-non-PT}(a)]. And the corresponding analytical solution is given in Appendix \ref{App1}, which proves that the perfect photon blockade cannot be achieved with the weak parameter mechanism when the two cavities are all passive. In addition, when the cavity 2 is an active cavity, the photon blockade phenomenon becomes more obvious for some certain detunings even with unbalance gain-loss ratio [see the blue circle line and the magenta triangle line in Fig.~\ref{fig:ga1-Delta1-non-PT}(a)]. Therefore, the balance gain-loss ratio is a necessary condition for the occurrence of the perfect photon blockade with the weak parameter mechanism.

On the other hand, the non-$\mathcal{PT}$ symmetric double-cavity optomechanical system can also be studied by setting $\Delta_{1}\neq\Delta_{2}$. The photon blockade phenomenon of the passive cavity 1 is studied and shown in Fig.~\ref{fig:ga1-Delta1-non-PT}(b) with different detunings of the active cavity 2. We find that the perfect photon blockade can also be achieved even when the detunings of the two cavities are different. However, the location of the perfect photon blockade is $\Delta_{1}=g^{2}/\omega_{m}-\Delta_{2}$ (see Appendix \ref{App2}), which is related to the optomechanical coupling strength, the mechanical frequency, and the detuning of the active cavity 2. The perfect photon blockade is located at $\Delta_{1}=0$ when the detuning of cavity 2 is chosen as $\Delta_{2}=\Delta_{1}+g^{2}/\omega_{m}$ [see the red square line in Fig.~\ref{fig:ga1-Delta1-non-PT}(b)]. Those results indicate that the main reason for the occurrence of the perfect photon blockade is the balance gain-loss ratio in the $\mathcal{PT}$-symmetric double-cavity optomechanical system and the different detunings of the two cavities only affect the location of the perfect photon blockade.

Finally, we briefly discuss the experimental feasibility of the present proposal. The $\mathcal{PT}$-symmetric coupled cavity systems have been widely studied in the experiment~\cite{Nature.488.167,NatMat.12.108,NatPhotonics.8.524,Science.346.975}. However, the single photon optomechanical coupling strength $g$ is still difficult to reach the coupling region ($g>\kappa_{1}$) for current experiments, e.g., $g\sim10^{-3}\kappa_{1}$ in photonic crystals~\cite{Nature.478.89} and microresonators~\cite{Nature.482.63}. Though the present proposal can achieve the photon blockade with $g\ll\omega_{m}$, which has relaxed the experimental requirements for the usual optomechanical systems~\cite{PhysRevLett.107.063601,PhysRevLett.107.063602,PhysRevA.87.025803,PhysRevA.88.023853,PhysRevA.92.033806,PhysRevA.93.063860,arXiv:1802.09254}, it still requires the optomechanical coupling is comparable to the cavity decay $g\sim\kappa_{1}$, which is still difficult for current experimental techniques. On the other hand, the effect of $\mathcal{PT}$ symmetry on photon blockade might be proved in the $\mathcal{PT}$-symmetric coupled system including a Kerr-type nonlinear medium, where the required Kerr-type nonlinear strength is just about $10^{-1}\kappa_{1}$ and can be implemented experimentally.

\section{Conclusions}\label{sec.5}
In conclusion, we have studied the photon blockade effects in $\mathcal{PT}$-symmetric double-cavity optomechanical system, where the optomechanical cavity is passive and the other one is an active cavity. Through the analytical solution and numerical simulation, we respectively obtain the equal-time second-order correlation functions, which are used to describe the photon blockade effects and agree with each other very well. We find an interesting phenomenon, where both of the cavities can be blocked at the same time when the detuning of the system equals to an appropriate value. Furthermore, we find that the photon blockade phenomenons are completely different in the broken and unbroken $\mathcal{PT}$-symmetric regions of the system. The CPB induced by the anharmonicity of the eigenenergy spectrum occurs only in the unbroken $\mathcal{PT}$-symmetric region. However, the UPB coming from interference between paths with gain and loss can occur whether the system is in the broken or unbroken $\mathcal{PT}$-symmetric region.

We strictly analyze and discuss the reason for the occurrence of the perfect photon blockade via the CPB and UPB theories, respectively, and find that the present UPB is slightly different from the usual UPB theory. Moreover, to further explain the photon blockade mechanism, we study the correlation function with the non-$\mathcal{PT}$ symmetric situations, e.g., unbalance gain-loss ratio or unequal detunings. Those results indicate that the perfect photon blockade can be achieved when the gain-loss ratio is balance and the unequal detunings just change the location of the perfect photon blockade. Compared with the traditional optomechanical system, the photon blockade in the $\mathcal{PT}$-symmetric double-cavity optomechanical system does not require the strong optomechanical coupling strength and can be achieved with the weak parameter mechanism, i.e., $g\ll\omega_{m}$ and $J<\kappa_{1}$. Therefore, our proposal might be more feasible in experiment and would contribute to the generation of the single photon source.

\begin{center}
{\bf{ACKNOWLEDGMENTS}}
\end{center}
This work was supported by the National Natural Science Foundation of China under Grant Nos. 61822114, 61465013, 61575055, and 11465020, and the Project of Jilin Science and Technology Development for Leading Talent of Science and Technology Innovation in Middle and Young and Team Project under Grant No. 20160519022JH.

\appendix
\section{double-passive-cavity optomechanical system}\label{App1}
Here, as a comparison, we study the photon blockade of the double-passive-cavity optomechanical system; namely, $\kappa_{2}=-\kappa_{1}$, where the system is non-$\mathcal{PT}$ symmetric. Similar to the calculation in Sec.~\ref{subsec.3A}, we can obtain the steady-state probability amplitudes
\begin{eqnarray}\label{Ae01}
C_{10}&=&\frac{2\omega_{m}E(2\Delta_{1}-i\kappa_{1})}{2g^{2}(2\Delta_{1}-i\kappa_{1})-\omega_{m}[(2\Delta_{1}-i\kappa_{1})^{2}-4J^{2}]},\cr\cr
C_{01}&=&\frac{-4J\omega_{m}E}{2g^{2}(2\Delta_{1}-i\kappa_{1})-\omega_{m}[(2\Delta_{1}-i\kappa_{1})^{2}-4J^{2}]},\cr\cr
C_{20}&=&2\sqrt{2}\omega_{m}^{2}E^{2}(2\Delta_{1}-i\kappa_{1})^{2}[g^{2}-\omega_{m}(2\Delta_{1}-i\kappa_{1})]/M,\cr\cr
C_{11}&=&8J\omega_{m}^{2}E^{2}(2\Delta_{1}-i\kappa_{1})[2g^{2}-\omega_{m}(2\Delta_{1}-i\kappa_{1})]/M,\cr\cr
C_{02}&=&8\sqrt{2}J^{2}\omega_{m}^{2}E^{2}[2g^{2}-\omega_{m}(2\Delta_{1}-i\kappa_{1})]/M,
\end{eqnarray}
with
\begin{eqnarray}\label{Ae02}
M&=&\{2g^{2}(2\Delta_{1}-i\kappa_{1})-\omega_{m}[(2\Delta_{1}-i\kappa_{1})^{2}-4J^{2}]\}\cr\cr
&&\times\{4g^{4}(2\Delta_{1}-i\kappa_{1})-g^{2}\omega_{m}[5(2\Delta_{1}-i\kappa_{1})^{2}-8J^{2}]\cr\cr
&&+\omega_{m}^{2}(2\Delta_{1}-i\kappa_{1})[(2\Delta_{1}-i\kappa_{1})^{2}-4J^{2}]\}.
\end{eqnarray}

To achieve the photon blockade of the cavity 1, we need to choose a set of system parameters to make $C_{20}=0$. That is to say, the system parameters should satisfy the follow relations
\begin{eqnarray}\label{Ae03}
\begin{cases}
12\Delta_{1}^{2}-4\Delta_{1}\frac{g^{2}}{\omega_{m}}=\kappa_{1}^{2}>0,\\
4\Delta_{1}(g^{2}-4\Delta_{1}\omega_{m})^{2}=0.
\end{cases}
\end{eqnarray}
It is easy to find that the above relations cannot be satisfied at the same time. That is to say that the perfect photon blockade of the cavity 1 cannot be achieved with the weak parameter mechanism in the double-passive-cavity optomechanical system. In addition, the parameter conditions of $C_{02}=0$ and $C_{11}=0$ are also unsatisfed due to $J\neq0$. So we can conclude that the perfect photon blockade cannot be achieved in the current situation. However, when the optomechanical coupling strength is enhanced~\cite{PhysRevA.87.013839} or the location of the driving field is changed~\cite{JPB.46.035502}, it will become possible to achieve the perfect photon blockade with the strong system parameters. Here, we would not discuss them in detail.

\section{unequal detunings}\label{App2}
For the non-$\mathcal{PT}$ symmetric system with different detunings, $\Delta_{1}\neq\Delta_{2}$, we can also obtain the probability amplitudes via solving the  Schr\"{o}dinger equation. Here, the probability amplitude $C_{20}$ is
\begin{eqnarray}\label{Be01}
C_{20}&=&2\sqrt{2}\omega_{m}^{2}E^{2}(2\Delta_{2}+i\kappa_{1})^{2}[g^{2}-\omega_{m}(\Delta_{1}+\Delta_{2})]/\cr\cr
&&\{[4J^{2}\omega_{m}+(2\Delta_{2}+i\kappa_{1})(2g^{2}-\omega_{m}(2\Delta_{1}-i\kappa_{1}))]\cr\cr
&&\times[4J^{2}\omega_{m}(2g^{2}-\omega_{m}(\Delta_{1}+\Delta_{2}))+(2\Delta_{2}+i\kappa_{1})\cr\cr
&&\times(g^{2}-\omega_{m}(\Delta_{1}+\Delta_{2}))(4g^{2}-\omega_{m}(2\Delta_{1}-i\kappa_{1}))]\},\cr\cr
&&
\end{eqnarray}
where $C_{20}=0$ can be obtained when $\Delta_{1}+\Delta_{2}=g^{2}/\omega_{m}$; namely, the perfect photon blockade is located at the point of $\Delta_{1}=g^{2}/\omega_{m}-\Delta_{2}$. That means that the perfect photon blockade can be achieved even when the system is non-$\mathcal{PT}$ symmetry and the location of the perfect photon blockade is related to the system detunings.

\end{document}